# Detection of fraudulent users in P2P financial market

*Hao* Wang*

HC Research, HC Financial Service Group, China

**Abstract.** Financial fraud detection is one of the core technological assets of Fintech companies. It saves tens of millions of money from Chinese Fintech companies since the bad loan rate is more than 10%. HC Financial Service Group is the 3$^{rd}$ largest company in the Chinese P2P financial market. In this paper we illustrate how we tackle the fraud detection problem at HC Financial. We utilize two powerful workhorses in the machine learning field - random forest and gradient boosting decision tree to detect fraudulent users. We demonstrate that by carefully select features and tune model parameters, we could effectively filter out fraudulent users in the P2P market.

## 1 Introduction

Fintech is one of the most thriving industry in many countries over the world. People have been relying on Fintech to lend and borrow money , detect fraudulent users , match loans between money lenders and borrowers. P2P is a business model where money lenders can distribute his investment over multiple borrowers and each borrower could gather money from a host of money lenders. P2P has been extremely popular in China with an annual interest return rate of 8% - 10% satisfying to most of P2P sites' users.

Major P2P companies in China have been using a technology called knowledge graph to facilitate their financial processes. Knowledge graph is a graph constructed from data collected from the internet with user's authorization, including the user's phone log , ID card information, bank account transactions, home addresses, etc. Each node in knowledge graph represents entity such as person, id card number, address etc. , while each edge represents relationship such as colleagues, family membership, etc. Knowledge graph is one of the core assets of P2P companies because it is highly useful in crucial business processes within the company. For example, knowledge graph could be used to do anti-fraud , which saves tens of millions of money given the bad loan rate is close to 50% on most of the P2P platforms.

HC Financial Service Group is one of the top P2P companies in China. We have a fully built team consisting of nearly 100 staff working on credit risk modeling problems taking advantage of more than 400 million users' information. Credit risk modeling problems , when solved in machine learning contexts, are conventionally modeled as class imbalance problems. Most of the time , feature engineering and graph pattern analysis play a key role in the problem solving process. Case-by-case and manual inspection of individual node and its neighbors in extremely large knowledge graph is daily routine of credit risk modeling team's work.

---

* Corresponding author: haow85@live.com

One advantage of the problem setting in P2P financial market is that the fraud rate is very high - as high as more than 10% , some times a lot higher. The high fraud rate causes big headache for company runners but saves the day for algorithm engineers since class imbalance problem is a lot less severe. In this paper, we demonstrate that using random forest and gradient boosting decision tree, we could obtain evaluation metrics comparable to non-class imbalance problems.

## 2 Related work

Fraudulent behavior exists since the advent of internet. Special groups and teams of internet companies and educational institutes are formed to tackle various fraudulent behavior. Facebook designed an algorithm called CopyCatch [1] to detect the Lockstep behavior. Later they came up with a new invention called SynchroTrap [2] to detect synchronized attack. CMU researchers invented fBox [3] and SpokeEigen [4] algorithms to detect community fraud.

Fraud detection is one of the most extensively researched field in Fintech industry. Common fraud detection methods include rule-based approaches, Bayesian networks and machine learning techniques. Fraud detection, when formulated as a classification problem, is inherently a class imbalance problem very difficult to solve. Therefore in practice expert domain knowledge has become one of the most important ingredients of the Fintech fraud detection system.

Vlasselaer et al. [5] provided a machine learning framework for graph-based financial fraud detection in general. They pointed out that graph pattern mining alone is seldom used as a standalone model for financial fraud detection. Graph patterns usually serve as features to classification models such as random Logistic regression.

## 3 Algorithm

Feature engineering is a crucial step in credit risk modeling . Selection and processing of the appropriate features is a delicate art. At HC Financial Group, we have both online and offline systems where we could store detailed information about money borrowers. For example, at our outlet in Beijing, our sales staff help customers input their information into our systems. Once the information is entered, our online system will ask the user to authorize us to acquire other information about the user such as the credit report from the People's Bank of China. We have hundreds of data items we could use for each user in our systems.

However, not every data item store in our system is necessary in our credit risk modeling processes. In our problem context, we select the following features as input to our models:

1. Financial information : Features in this group include user's personal income , car rent , house rent , etc.

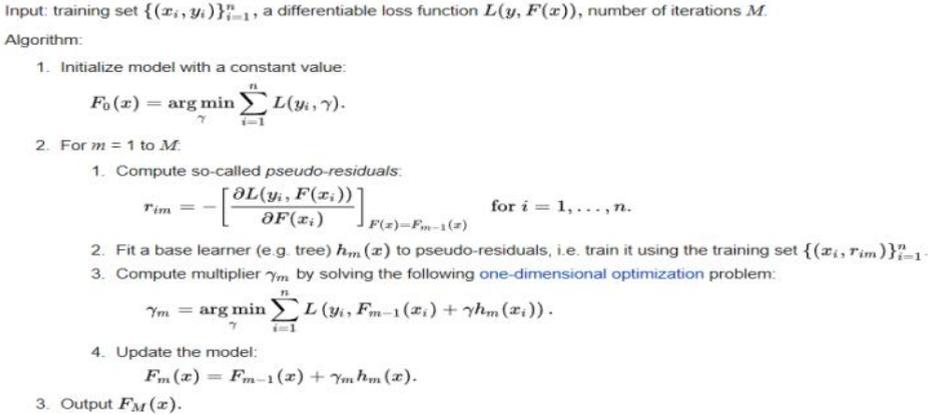

Figure 1 General Work Flow of Gradient Boosting Algorithms

2. Work information : Features in this group include user's company's income , how long the company has been founded, etc.

3. Transaction information : Features in this group include the amount of money user borrows in the transaction , whether the user has submitted applications before , etc.

4. Demographic information: Features in this group include the number of family members of the user, etc.

We chose these 4 groups of features because they are both heuristically and empirically consistent with our mission. For categorical data in the features, we utilize one-hot encodings to expand the term into several terms consisting of either 1 or 0. There are 97 selected features in total with 33 being categorical features.

To compare different feature engineering results, we use original data , data with PCA dimensionality reduction , data with tanh conversion , among many other feature engineering schemas. We would like to try PCA because after one-hot encoding, the feature space is expanded into hundreds of dimensions. We explore tanh because of the sharp contrast in the magnitudes of different features.

We use two of the most powerful workhorses in the machine learning field to detect the fraud users - random forest and gradient boosting decision tree. We choose these two models because they provide easy-to-tune parameters and robust results.

Random forest is an ensemble learning algorithm that aggregates the results from a group of regression or classification trees. The parameters that needs to be tuned include the maximum depth of each tree and the total number of trees in the forest.

Gradient boosting decision tree (GBDT) is one of the most successful gradient boosting algorithmic paradigms. The general work flow of a gradient boosting algorithm [6] is listed in Figure 1. GBDT utilizes the general gradient boosting schema with trees as its elementary components. The parameters that needs to be tuned include the maximum depth of each boosting tree, learning rate of the algorithm , number of trees in the model, etc.

To evaluate our algorithms, we use the AUC metric. We choose AUC because it is insensitive to class balance ratio. AUC measures the area of the ROC curve. A random classification result is close to 0.5. A perfect classification result is close to 1.0. The higher the value of the AUC, the better the classification result is.

The major toolkit we use to develop and test our algorithms is scikit-learn and xgboost.

## 4 Experiment

We select two datasets to test our algorithms : One consists of 30K normal users and 30K users with overdue payments (dataset A) , the other consists of 50K users consisting of 25K fraud users and 25K normal users (dataset B) . We split our dataset into training set, test set and validation set with the ratio of 4:1:1.

In our random forest algorithm, we enumerate the values of the maximum depth of each tree and the number of trees in the model. Figure 1 illustrates the scatter plot visualizing the results (AUC) of random forest + PCA dataset A. Maximum depths of trees are enumerated between 2 and 5 while the number of trees in the forest is enumerated between 5 and 120 . As shown in the figure, AUC increases with max depths until it reaches the value of 4. It is comparatively insensitive to the number of trees in the forest.

Figure 2 visualizes the change of parameters with function tanh applied to features so they become normalized. Tanh generates better results than PCA only or PCA and tanh combined , with the average AUC being 0.780 on test set and 0.797 on validation set.

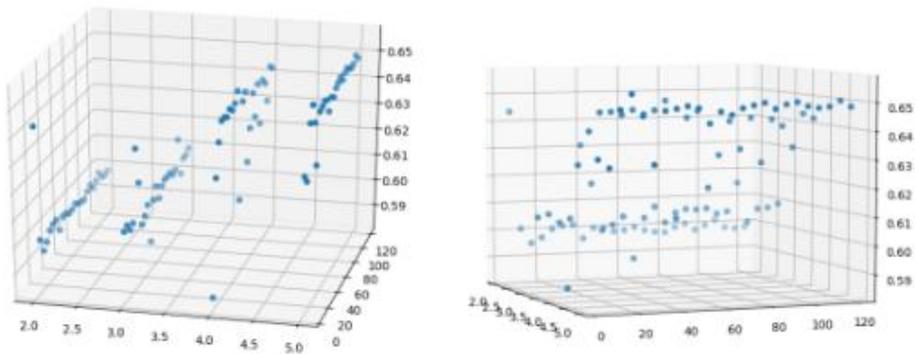

**Fig. 1.** Scatter plot of random forest + PCA on dataset A.

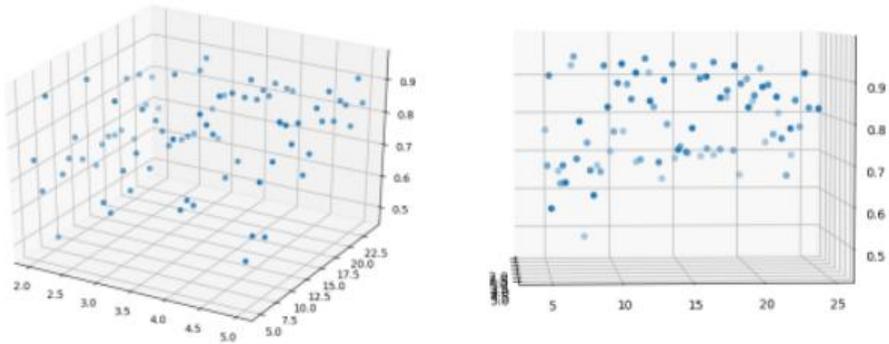

**Fig. 2.** Scatter plot of random forest + tanh on dataset A.

We also test random forest on dataset B. We enumerate the maximum depths and number of trees as the parameters to the random forest algorithm. Figure 3 shows the result with tanh function applied to the features . Similar to our result on dataset A, tanh alone generates better results than PCA or PCA and tanh combined.The average AUC metrics on both test and validation sets are 0.83.

Figure 4 shows the results of tanh applied to GBDT input feautres on dataset B. After having eliminated a few AUC outliers, the average AUC on both test and validation sets yields values close to 0.88.

In both data sets, PCA generates more consistent AUC metric per each parameter compared to the same algorithm on plain data set. However , tanh operation on features yield better scores than PCA only or PCA / tanh combined on both data sets. We see from our experimental results, partially due to the nice data structure, simple tweaks after careful feature selection could lead to results usable for our online systems.

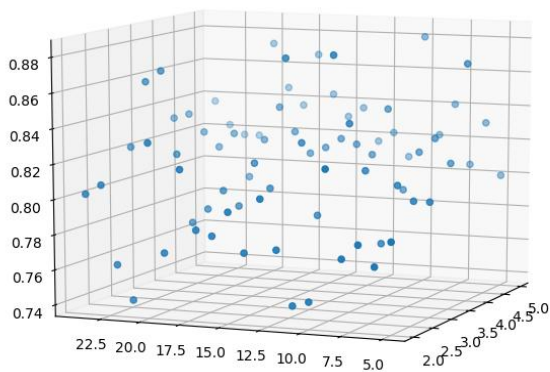

**Fig. 3.** Scatter plot of random forest + tanh on dataset B.

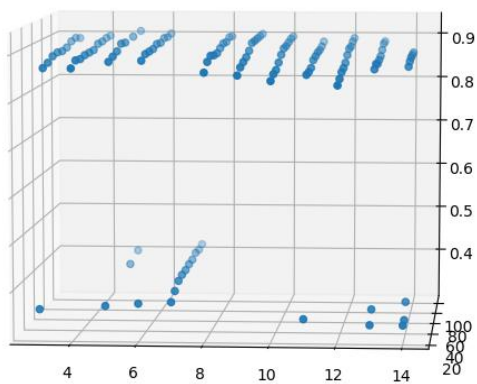

**Fig. 4.** Scatter plot of GBDT + tanh on dataset B.

## 5 Conclusion

Fraud and overdue payment users detection is a crucial step in Fintech company's business processes. Effective filter-out of such users could save tens of millions of US dollars for the company. In this paper we propose using feature engineering and random forest / GBDT algorithms to detect fraudulent users in P2P financial market. Due to the high fraud rate in the market (more than 10%), it is comparatively easy to generate satisfying results compared to other markets like conventional banking systems where the fraud rate could be as low as 2%. We tested our algorithms on two real world data sets and visualized the AUC metric together with the selection of model parameters.

In future work, we would like to explore the possibility of taking advantage of deep learning techniques to detect fraud and overdue payment users in our systems. Deep learning has been effectively and ubiquitously used in a whole spectrum of machine learning tasks. The expectation of better evaluation metric results could never be underestimated.